\documentclass[12pt]{article}
\usepackage{graphicx}

\newsavebox{\sboxpubnumber}
\newsavebox{\sboxpubdate}
\newcommand{\pubdate}[1]{\begin{lrbox}{\sboxpubdate}{#1}\end{lrbox}}

\newcommand{\Title}[1]{\begin{center} {\Large #1 } \end{center}}
\newcommand{\Author}[1]{\begin{center}{ \sc #1} \end{center}}
\newcommand{\Address}[1]{\begin{center}{ \it #1} \end{center}}
\newcommand{\andauth}{\begin{center}{and} \end{center}}

\newenvironment{Abstract}{\begin{quotation}  }{\end{quotation}}
\newenvironment{Presented}{\begin{quotation} \begin{center}
             PRESENTED AT\end{center}\bigskip
      \begin{center}\begin{large}}{\end{large}\end{center}
      \end{quotation}}
\newcommand{\Acknowledgements}{\bigskip  \bigskip \begin{center} \begin{large}
             \bf ACKNOWLEDGEMENTS \end{large}\end{center}}

\def\gsim{\;\raise0.3ex\hbox{$>$\kern-0.75em\raise-1.1ex\hbox{$\sim$}}\;}
\def\lsim{\;\raise0.3ex\hbox{$<$\kern-0.75em\raise-1.1ex\hbox{$\sim$}}\;}

\begin{document}

\begin{titlepage}
\pubdate{\today}                    

\vfill
\Title{Nonthermal production of gravitinos and inflatinos in the Inflationary Universe}
\vfill
\Author{Hans Peter Nilles}

\vfill
\andauth
\vfill
\Author{Marco Peloso}
\Address{Physikalisches Institut, Universit{\"{a}}t Bonn,\\
Nussallee 12, D-53115 Bonn, Germany}

\vfill \begin{Abstract}

The success of primordial nucleosynthesis imposes stringent bounds on the
abundance of gravitational relics. This is particularly true for
gravitinos, which - for models with gravitationally mediated supersymmetry
breaking - are expected to have a mass below the TeV scale and thus to
decay only after nucleosynthesis has concluded. We discuss the nonthermal
production of gravitinos in models with several chiral fields, to be able
to distinguish the mechanisms of supersymmetry breaking and inflation. Our
explicit calculations show that the superpartner of the inflaton can be
significantly generated. Gravitinos can be produced both at the preheating
stage and at the subsequent inflaton/inflatino decay. We verify that this
production is well below the nucleosynthesis bound, provided the sector
responsible for the present supersymmetry breakdown is weakly (for
example, only gravitationally) coupled to the one which drove inflation.

\end{Abstract}

\vfill
\begin{Presented}
    COSMO-01 \\
    Rovaniemi, Finland, \\
    August 29 -- September 4, 2001
\end{Presented}
\vfill
\end{titlepage}
\def\thefootnote{\fnsymbol{footnote}}
\setcounter{footnote}{0}

\section{Introduction}

There are several issues in astrophysics and cosmology where very high
energy scales seem to be involved. For instance, consider the observation
of cosmic rays with energy above the Greisen-Zatsepin-Kuzmin  cut-off, or
baryogenesis schemes as leptogenesis and GUT baryogenesis. The former may
be explained with the decay of very massive particles with a lifetime
comparable or larger than the present age of the Universe; the latter
involve very massive right handed neutrinos or GUT bosons, whose out-of
equilibrium decay generates the tiny baryonic asymmetry suggested by
primordial nucleosynthesis. In most cases, these heavy particles have
masses higher than the one of the inflaton, which  typically ranges in the
interval $\left( 10^{10} - 10^{13} \right)$ GeV in the simplest models of
``new'' and ``chaotic'' inflation. As a consequence, reheating based on a
perturbative decay of the inflaton can hardly account for their
generation. Fortunately, the last decade has witnessed a drastic change in
the theory of reheating, with the realization that nonperturbative effects
can play a very significant role. The nonperturbative decay of the
inflaton is typically a very quick phenomenon, which is expected to be
followed by a phase of ordinary reheating or thermalization. For this
reason, it is commonly denoted as {\it preheating}. Preheating of bosons
is characterized by a very efficient and explosive creation, due to the
coherent effect of the oscillations of the inflaton field. This allows
significant production even when single particle decay is kinematically
forbidden. It has been successively realized that preheating of fermions
can also be very efficient despite of the fact that the parametric
resonance is in this case limited by Pauli blocking.

One of the main successes of inflation is the generation of a nearly scale
invariant spectrum of primordial metric fluctuations, as indicated by
Cosmic Microwave Background and Large Scale Structure measurements. While
the nearly scale invariance is a natural outcome of the slow roll of the
inflaton field, the observed smallness of perturbations typically requires
an {\it ad hoc} fine tuning of some parameter of the inflationary sector.
As an example, in the two most standard models of single field chaotic
inflation, $V \left( \phi \right) = m_\phi^2 \, \phi^2/2$ and $V \left(
\phi \right) = \lambda \phi^4 \, /4\,$, one has to fix $m_\phi^2 \sim
10^{-\,11} \, M_P^2$ for a massive inflaton ($M_P$ here denotes the
reduced Planck mass) or $\lambda \sim 10^{-\,13}$ in the massless inflaton
case. Supersymmetry may provide a natural framework for the protection of
such small couplings, and indeed inflationary supersymmetric models have
been widely discussed in the literature. Since in chaotic and new
inflationary schemes the inflaton field acquires values comparable or
larger than $M_P\,$, complete models should include also supergravity
effects rather than just global supersymmetry. These effects are mostly
relevant at very early times, but one has to check that they are
consistent with the phenomenology of the later evolution of the Universe.
The most known example of possible difficulties in this direction is
certainly provided by the {\it gravitino problem}.

The gravitino is necessary present in supergravity, since it is the
supersymmetric partner of the graviton. It has four physical degrees of
freedom, two ``transverse'' and ``two'' longitudinal. Actually, the
longitudinal components are present only for broken supersymmetry, and
they are provided by the goldstino field, which is a linear combination of
the partners of the scalars (or gauge bosons) responsible for the
supersymmetry breakdown. This mechanism is the ``fermionic version'' of
spontaneous breaking of gauge symmetries, and indeed it is known as
superhiggs mechanism. If supersymmetry has to solve the hierarchy problem,
the scale of its breaking, and thus the mass of the gravitino (we restrict
here to models of gravitational mediated supersymmetry breaking) cannot be
too far from the electroweak scale, $m_{3/2} \sim 100 \, {\rm GeV} - 1 \,
{\rm TeV}$. If not the lightest supersymmetric particle, a gravitino with
a mass in this range will decay only after nucleosynthesis has ended. The
decay typically produces an electromagnetic shower, which can alter the
successful predictions of standard primordial nucleosynthesis. Indeed, not
to conflict with observations, the very strong limit
\begin{equation}
Y_{3/2} \equiv \frac{n_{3/2}}{s} \lsim 10^{-\,13} \,\,,
\label{limit}
\end{equation}
has to be imposed on the gravitino abundance (in standard notation, here 
$n_{3/2}$ denotes the gravitino number density, while $s$ the entropy density).

In the last twenty years, detailed calculations have been performed on the
perturbative production of gravitinos after the inflaton decay products
have thermalized. One finds that, for a gravitino with mass in the above range, a rather strong bound on the reheating
temperature,
\begin{equation}
T_{\rm rh} \lsim \left( 10^{9} - 10^{10} \right) \; {\rm GeV} \,\, ,
\label{temp}
\end{equation}
has to be imposed in order to avoid excessive production. More recently,
it has been wondered if preheating can also lead to a too large creation
of gravitinos. This issue is, however, more complicated. Quite generally,
preheating effects lead to the non-perturbative production of the
fermionic partner of the inflaton, the inflatino, and of any other fermion
which is strongly coupled to the inflaton field. While one easily finds
that the transverse gravitino component is only weakly (i.e.
gravitationally) coupled to the inflaton, and that its quanta are hence
produced in a negligible amount, the production of the longitudinal
component is very model dependent. If there is substantial mixing between
the inflatino and the longitudinal component of the gravitino, the
goldstino, preheating may result in an overproduction of gravitinos.
During inflation and the beginning of reheating, supersymmetry is mainly
broken by the inflaton implying a strong correspondence between the
inflatino and goldstino at this early stage.  However, this correspondence
does not necessarily hold at late times, since supersymmetry may be 
broken by other fields in the true vacuum of the theory. Indeed, this is
probably the most typical situation, as it is natural to distinguish
between inflation and supersymmetry breaking due to the very different
energy scales associated with the two phenomena. In this case, the final
gravitino abundance can be much smaller than the inflatino one.

More accurately, the relic abundance of gravitinos will ultimately be
related to the strength of the coupling between the inflationary sector
and the one responsible for the present supersymmetry breakdown. The
simplest possibility is to consider a model with two fields coupled only
gravitationally, which is a simple prototype of more realistic schemes of
gravitational mediation of supersymmetry breaking in a hidden sector. Even
in this simplified model, the calculation of nonthermal gravitino
production requires substantial work. First, one has to develop a new
formalism, to be able to clearly define and compute the production in
systems with several coupled fields. Subsequently, an extended numerical
investigation has to be carried out to obtain reliable results. This
calculation, which we have described in details in the works~\cite{uno,
due}, is here summarized in section~\ref{sec1}.

In accordance with the above discussion, the final gravitino production in
these schemes turns out to be completely negligible also for the
longitudinal component. The nonthermal production is much more relevant
for the fermionic partner of the inflaton, the inflatino, which in some
cases can be produced with a significant abundance. Thus, gravitinos may
be overproduced through inflatino decay, if the channel inflatino
$\rightarrow$ inflaton (or its scalar partner) $+$ gravitino is
kinematically allowed. Depending on the relative masses of the inflaton
and inflatino, significant production may be expected instead by the
inverse process. This production may be particularly significant if the
inflaton sector is coupled only gravitationally to matter, since in this
case the above decays will have a rate comparable to the one generating
the thermal bath. Indeed, in such a scenario,  the decay channels into
gravitinos need to be strongly suppressed. In the work~\cite{tre} we have
discussed the possible kinematic suppression of this channel, which
translates into a rather strong lower bound on the scale of inflation. We
have shown that this bound is satisfied by the simple single scale
(supergravity) inflationary models. This analysis is summarized in
section~\ref{sec2}.

An extensive list of relevant bibliography can be found in~\cite{uno,due,tre}.

\section{Gravitino production at preheating} \label{sec1}

The system we are considering has the matter content of two superfields
$\Phi$ and $S$, with superpotential
\begin{equation}
W=\frac{m_\phi}{2}\,\Phi^2+\mu^2\,\left(\beta+S\right)
\label{sup}
\end{equation}
and with minimal K\"ahler potential $G = {\cal {K}} + {\rm ln} \vert W
\vert^2 \;\;,\;\; {\cal {K}}=\Phi^\dagger\, \Phi+S^\dagger\, S$. The field
$\phi$ (that is, the scalar component of $\Phi\,$) acts as the inflaton.
In the present model, supergravity corrections spoil the flatness of the
potential during inflation, so that additional contributions must be
relevant during inflation. However, we are interested in the dynamics of
the system {\it after} inflation, when  $\langle \phi \rangle \lsim M_p$,
and supergravity corrections are not important. We then assume that the
superpotential~(\ref{sup}) is valid at this stage, but we nonetheless
normalize $m_\phi\sim 10^{13}$ ~GeV, as required by the COBE normalization
of the CMB fluctuations for the ``usual'' chaotic inflation. A model where
also new inflation is obtained was considered in~\cite{tre}, and it gives
the same qualitative results as the one here described.

The superfield $S$ leads to the breaking of supersymmetry in the true
vacuum owing to its ``Polonyi'' superpotential. By imposing
$\beta=\left(2-\sqrt{3}\right)\,M_p$, one can indeed break supersymmetry
while retaining a vanishing cosmological constant in the true vacuum,
where the fields $s$ (the scalar component of the superfield $S$) has
expectation value of the order of $M_p$. The gravitino mass in the vacuum
is of the order $\mu^2/M_p$. In order to have a gravitino mass of about
$100$~GeV (that is the expected value for the gravitino mass in
gravity--mediated supersymmetry breaking models), $\mu \sim 10^{10}$~GeV
is required.

Right after inflation the field $\phi$ is oscillating about the bottom of its potential with frequency proportional to $m_\phi$. The time dependent expectation value of $\phi$ acts as an effective mass for the Polonyi scalar, which has vanishing expectation value at this stage. In this initial period the (time dependent) expectation value of $\phi$ is the main source of supersymmetry breaking. The amplitude of the oscillations of the field $\phi$ eventually decreases, due to the expansion of the Universe, and for times of the order of $m_{3/2}^{-1}$ the Polonyi scalar starts rolling down towards its true minimum and then oscillates about it.~\footnote{We neglect here the Polonyi problem associated with
the late times oscillations of $S\,$.} 

The system is thus governed by two time scales. At ``early'' times, of the
order of $m_\phi^{-1}$, the only relevant dynamics is the one of the
inflaton sector, that is also the main source of supersymmetry breaking.
At ``late'' times, much larger than $m_{3/2}^{-1}$, the system behaves as
if it was in its true vacuum, and supersymmetry is broken by the Polonyi
sector.  To be more specific, we define the dimensionless parameter
$\hat{\mu}^2 \equiv \mu^2/ \left( m_\phi\,M_p \right)\sim m_{3/2}/m_\phi$,
that gives the ratio of the two time scales in the system.  If
supersymmetry is supposed to solve the hierarchy problem, $\hat{\mu}^2$
should be of the order of $10^{-11}$. Such a small parameter implies a
very large difference between the two time scales of the problem, which is
a source of technical difficulties in the numerical computations. As a
consequence, we could not study the evolution of the system for such a
small value of $\hat{\mu}^2$.  Thus, we kept it as a free parameter and we
studied how a variation of $\hat{\mu}^2$ affects the scaling of the
relevant quantities. 

During the evolution, both the kinetic and potential energies of the two fields contribute to the supersymmetry breaking. One can define
\begin{equation}
f_{\phi_i}^2 \equiv m_i^2 + \frac{1}{2} \, \left( \frac{d \phi_i}{d t}
\right)^2 \;\;,
\end{equation}
with $m_i = {\rm exp} \left( {\cal {K}} \, M_p^{-\,2} / 2 \right) \,
\left[ \partial_i W + M_p^{-\,2} \, \partial_i \, {\cal {K}} \, W
\right]\,$. The quantities $f_i$ give a ``measure'' of the size of the
supersymmetry breaking provided by the $F$  term associated with the
$i-$th scalar field. More precisely, we will be interested in the
normalized quantities $r_i \equiv f_i^2 / \left( f_1^2 + f_2^2 \right)\,$,
which indicate the relative contribution of the two scalars. It can be
easily verified~\cite{uno} that in the initial stages $r_\phi \simeq 1\,$,
while $r_s \simeq 1$ at late times. The regime of equal contribution is 
around $t\, m_\phi={\hat \mu}^{-\,2}\,$.

Let us now consider the fermionic content. We denote the fermions of the
two chiral multiplets by ${\tilde \phi}$ (the ``inflatino'') and ${\tilde
s}$ (the ``Polonyino''). One linear combination of them is the goldstino
$\upsilon\,$, while the one orthogonal to $\upsilon\,$ is denoted by
$\Upsilon\,$. Initially, $\upsilon \equiv {\tilde \phi}\,$, while
$\upsilon \equiv {\tilde s}$ at late times. In addition, we have the
gravitino field, whose longitudinal and transverse component are denoted
by $\theta$ and by $\psi_i^T\,$, respectively. The transverse component is
decoupled from the other fermion fields, and its quanta are produced only
gravitationally. The longitudinal gravitino component, which in the
super-higgs mechanism is provided by the goldstino $\upsilon\,$, is
however coupled with $\Upsilon\,$. The computation of the occupation
numbers of the fermions $\theta$ and $\Upsilon$ is far from trivial and
requires a significant extension of the existing formalism for
nonperturbative production in the one field case. The details of the
calculation are reported in~\cite{due}. Here we will only present our
results.

\begin{figure}[htb]
\centering
\includegraphics[height=4in, angle=-90]{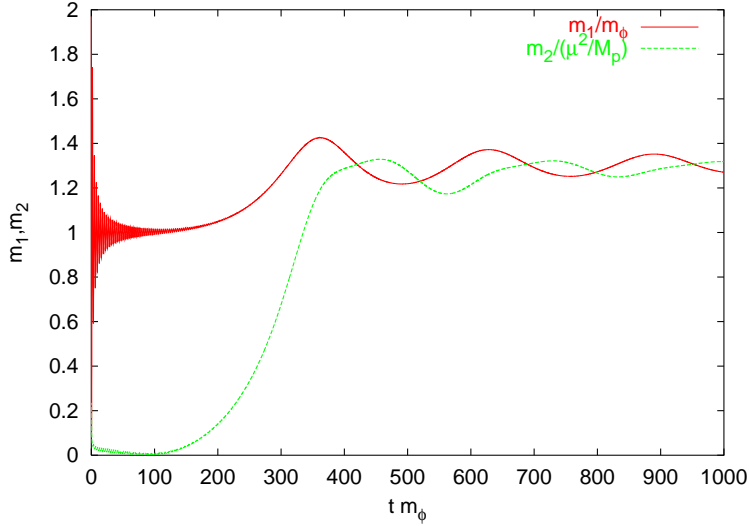}
\caption{Evolution of the masses of the two fermionic eigenstates. For illustrative porpoises we take ${\hat \mu}^2 = 10^{-\,2}\,$. Notice the different normalizations for the two masses.}
\label{fig1}
\end{figure}

As a starting point, one has to diagonalize (at each time) the coupled
$\theta$--$\Upsilon$ system. We denote the two fermionic mass eigenstates
by $\psi_1$ and $\psi_2\,$. In fig.~\ref{fig1} we show~\footnote{From
fig.~\ref{fig1}, one may be tempted to identify $\psi_1 \equiv {\tilde
\phi}$ and $\psi_2 \equiv {\tilde s}\,$. Although this identification is
rigorous only at the beginning and at the end of the evolution, it can be
used for an ``intuitive'' understanding of the system.} the evolution of
their masses for the specific case ${\hat \mu}^2 = 10^{-\,2}\,$. The most
relevant information which can be constructed by this evolution is very
clear: at late times the fields $\psi_1$ and $\psi_2$ have, respectively,
the mass of the inflatino and of the gravitino field. That is, at late
times we have the identification $\psi_1 \equiv {\tilde \phi} \equiv
\Upsilon \;,\; \psi_2 \equiv \theta\;$ (${\tilde s} = \upsilon = 0 \,$,
being the goldstino). This situation is orthogonal to the initial one,
when $\psi_1 \equiv \theta \;,\; \psi_2 \equiv {\tilde s} \equiv \Upsilon
\;$ (${\tilde \phi} = \upsilon = 0$). At intermediate times, when
supersymmetry is broken by both scalar fields, the gravitino is a mixture
of $\psi_1$ and $\psi_2\,$. To qualitatively appreciate the evolution of
the gravitino occupation number, we may consider $N_\theta \equiv r_1 \,
N_1 + r_2 \, N_2$ and the orthogonal combination $N_\Upsilon \equiv r_2 \,
N_1 + r_1 \, N_2\,$, where $r_i$ are the relative contributions of the two
scalars to supersymmetry breaking defined above. 

\begin{figure}[htb]
\centering
\includegraphics[height=4in, angle=-90]{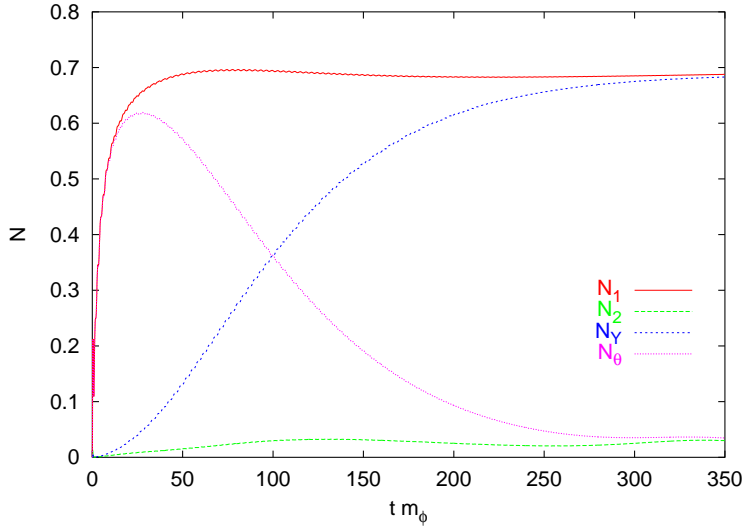}
\caption{Evolution of $N_\theta$ and $N_\Upsilon$ for ${\hat \mu}^2 =
10^{-\,2}\,$ and $k = m_\phi\,$. See the text for details.}
\label{fig2}
\end{figure}

We show in fig.~\ref{fig2} the evolution of $N_1 \,,\, N_2 \,,\, N_\theta
\,,\, N_\Upsilon$ for modes of comoving momentum $k=m_\phi$ (the scale
factor of the Universe is normalized to one at the end of inflation) and
for  ${\hat \mu}^2 = 10^{-\,2}\,$ . Notice that (by construction)
$N_\theta \equiv N_1$ at early times, while $N_\theta \equiv N_2$ at late
ones. We also see that $\psi_1$ is populated on time scales
$m_\phi^{-\,1}\,$, while $\psi_2$ on time scales ${\hat \mu}^{-\,2} \,
m_\phi^{-\,1}\,$. This feature is common for all ${\hat
\mu}^2\,$~\cite{due}. We remark that the identification $\theta \equiv r_1
\, \psi_1 + r_2 \, \psi_2$ should be taken only as a qualitative
indication. However, the most relevant identification $\theta \equiv
\psi_2$ at late times is a rigorous one, as should be clear from the above
discussion.

We are now ready to present our most important result: the occupation
number of the two fermionic mass eigenstates at the end of the process. 

As we have said, the realistic case ${\hat \mu}^2 = 10^{-\,11}$ is far
from our available resources, so we kept $\hat\mu^2$ as a free parameter
and performed numerical calculations down to ${\hat \mu}^2 = 10^{-\,6}\,$.
The results for ${\hat \mu}^2 = 10^{-\,11}$ can be clearly extrapolated
from the ones we are going to present. Moreover, the case $\hat{\mu}^2=0$
can be studied analytically~\cite{due}, and it agrees with the limit
${\hat \mu} \rightarrow 0$ that one deduces from the numerical results.
For ${\hat \mu} = 0\,$, only the inflatino is produced. The mass of the
Polonyi fermion does instead vanish identically, so no quanta of this
particle are produced at preheating~\cite{due}. Notice that ${\hat \mu} =
0$ corresponds to a situation with unbroken susy in the vacuum, and it
reproduces the models with one single field studied so far. We stress that
in this case only the inflatino is produced at preheating.

\begin{figure}[htb]
\centering
\includegraphics[height=4in, angle=-90]{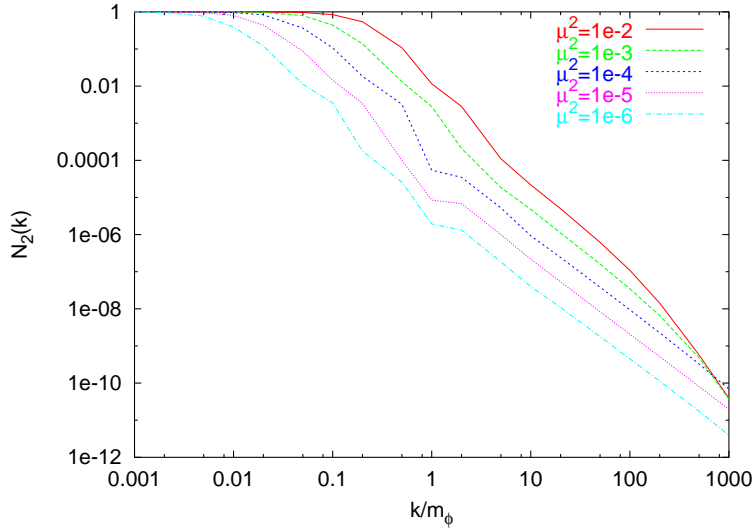}
\caption{Spectrum of gravitinos at late times.}
\label{fig3}
\end{figure}

As we showed in~\cite{uno,due}, the field $\psi_1$ has occupation number
of order one up to $k \sim m_\phi\,$, and then rapidly decreases as
$k^{-\,4}\,$. The final spectrum is practically independent of the value
of $\hat\mu^{-2}$. These features are easily explained: the eigenstate
$\psi_1$ is associated to the inflatino, that is produced by the coherent
oscillations of the inflaton; the inflaton dynamics occurs on time scales
of the order of $m_\phi^{-1}$, and is independent of $\hat\mu^2$. 

The spectrum of $\psi_2$ -- the late time gravitino -- is shown in
fig.~\ref{fig3} for different values of ${\hat \mu}^2\,$,~\footnote{These
spectra are shown at the time $t=10 \, {\hat\mu}^{-\,2} m_\phi^{-\,1}\,$.
In the ${\hat \mu} = 10^{-\,2} - 10^{-\,4}$ cases we have continued the
evolution further, until the spectra stop evolving. In these cases, we
have found that the spectra very slightly {\em decrease} for $t > 10 \,
{\hat\mu}^{-\,2} m_\phi^{-\,1}\,$. Thus, the results shown in
fig.~\ref{fig3} give an accurate upper bound on the final gravitino
abundance.} and exhibits instead a more significant dependence on
$\hat\mu^2$. One finds that the production of these quanta mainly occurs
during the first oscillations of the Polonyi field, with a typical physical
momentum of the order the Polonyi ($\sim m_{3/2}$) mass $\mu^2/M_P\,$. The
total number of particles produced in this case is thus an increasing
function of $\hat\mu^2$. One can easily realize that the production is
expected to be completely negligible in the physically relevant case
${\hat \mu}^2 = 10^{-\,11}\,$.

\section{Gravitino production through the inflaton/inflatino decay} \label{sec2}

In this section we generalize the results discussed above, considering
also gravitino production from the decay of inflaton and inflatino. Since
we will always consider a supersymmetry breaking scale much smaller than
the scale of inflation, it is worth noticing that the decays of the
inflaton and the inflatino are expected to occur nearly simultaneously. As
we have said, the potential of the inflaton field during inflation is
constrained by the magnitude of density fluctuations following from CMB
temperature measurements. In the slow roll regime, one has to impose
\begin{equation}
V_{60}^{1/4} \simeq 0.027 \: \epsilon_{60}^{1/4} \: M_P \,\,,
\label{cobeve}
\end{equation}
where $\epsilon^2 \equiv  M_p^2 \, \left( V'/V \right)^2 / 2 \ll 1$ is one
of the two slow-roll parameters (prime denoting derivative with respect to
$\phi$) and the suffix $60$ reminds us that the two quantities have to be
evaluated when the scales measured by COBE left the horizon, about $60$
e--foldings before the end of inflation. As stated in the introduction,
this relation uniquely fixes the scale of the potential in the simplest
models of one field inflation and potential $V \propto \phi^n\,$. However,
models with a much smaller scale than the one reported there and
acceptable density fluctuations can be constructed. As follows from
eq.~(\ref{cobeve}), in models with one single field this can be done at
the expense of a small $\epsilon$ parameter, that is by taking a very flat
potential during inflation. Such a flat potential may arise more naturally
if more scalar fields are present, as for example in hybrid inflationary
models. 

Due to this freedom, in this section we discuss the production of
gravitinos with a generic inflationary scale. To be precise, we denote by
$\Delta^4$ the value of the inflaton potential at the end of inflation,
when the reheating stage begins. Due to the slow motion of $\phi$ during
inflation, this scale is typically very close to $V_{60}\,$. During
reheating, the inflaton field oscillates about the minimum $\phi_0$  of
$V$, with $V \left( \phi_0 \right) = 0\,$. As is typical for a massive
inflaton, we assume that the quadratic term dominates the Taylor expansion
of $V$ around $\phi_0\,$.  We denote by $F$ the amplitude of the inflaton
oscillations at the initial time $t \sim H^{-\,1} \sim M_P / \Delta^2\,$.
In the present discussion we keep also the value of $F$ as a free
parameter. As we will now show, the number density of produced gravitinos
is more closely related to the inflaton mass $m_\phi$ rather than to the
scale $\Delta\,$. For a quadratic potential and generic values of $\Delta$
and $F\,$, one has $m_\phi \simeq \Delta^2 / F\,$. Note that the model
considered in the previous section is characterized by $F \simeq M_P$ and
$m_\phi \simeq 10^{13}\, {\rm GeV}$.

We discuss here the case in which the inflaton decays only
gravitationally. A more general analysis where also nongravitational
inflaton decays are considered was performed in~\cite{tre}. The
gravitational decay rate of the inflaton is given by $\Gamma \simeq
m_\phi^3/M_p^2\,$, leading to the reheating temperature $T_{\rm rh} \simeq
\left( \Gamma \, M_P \right)^{1/2} \simeq \sqrt{m_\phi^3/M_P}\,$. The
thermal bound~(\ref{temp}) thus simply gives
\begin{equation}
m_\phi \lsim 10^{12} \, {\rm GeV} \,\,.
\label{fbound1}
\end{equation}

The inflaton number density at its decay can be estimated to be $n_\phi =
V \left( \phi \right) / m_\phi$ at $t \equiv \tau_\phi = \Gamma^{-\,1}\,$. The inflaton ``abundance'' at the decay time is thus approximatively given by
\begin{equation}
Y_\phi \simeq \frac{n_\phi}{\rho_\phi^{3/4}} \simeq \frac{T_R}{m_\phi}
\simeq \sqrt{\frac{m_\phi}{M_P}}
\label{finfla}
\end{equation}
The abundance of inflatinos produced nonthermally by the inflaton
oscillations is also easily evaluated, by remembering that inflatinos are produced at preheating up to physical momentum $m_\phi\,$, and with a physical number density of the order
\begin{equation}
n_{\tilde \phi} \simeq 10^{-\,2} \, m_\phi^3 \, \left[ a \left( t = m_\phi^{-\,1} \right) / a \left( t \right) \right]^3\,\,.
\end{equation}
The inflatino abundance at $t=\tau_\phi$ is thus
\begin{equation}
Y_{\tilde \phi} \simeq \frac{n_{\tilde \phi}}{\rho_\phi^{3/4}} \simeq 10^{-\,2} \, \frac{m_\phi \, T_R}{M_P^2} \simeq 10^{-\,2} \left( \frac{m_\phi}{M_P} \right)^{5/2}
\label{inflab}
\end{equation}

Also the abundance of gravitinos produced nonthermally becomes
smaller as $m_\phi$ decreases, so that it is always negligible. Indeed,
gravitinos will still be mainly produced at the time $t \sim
m_{3/2}^{-\,1}$ with a typical momentum $k \sim m_{3/2}^{-\,1}\,$,
independent of the value of $m_\phi\,$. However, a lighter inflaton
implies a longer lifetime $\tau_\phi\,$. Since the quantity
$n_{3/2}/\rho_\phi^{3/4}$ decreases in the time interval $m_{3/2} < t <
\tau_{\phi}\,$, lowering $m_\phi$ will thus decrease the final nonthermal
gravitino abundance.

We see that the inflaton abundance is always higher than the one of
inflatinos. Indeed, once the bound~(\ref{fbound1}) is respected, the
inflatino abundance is smaller than the limit $10^{-\,13}$ reported in
equation~(\ref{limit}). Even if all the inflatino quanta produce a
gravitino when they decay, this would not lead to an overproduction of
gravitinos (the inflatino abundance can be higher for a massless inflaton,
e. g. with potential $V \left( \phi \right) \propto \phi^n \,\,,\,\, n >
2\,$; see~\cite{due,tre} for details). Gravitino production can instead be
significant through the $\phi \rightarrow {\tilde \phi} \, {\tilde G}\,$
decay, if kinematically allowed.

Indeed, assuming that the branching fraction for this decay is $1/N$, from
eq.~(\ref{finfla}) we get the gravitino abundance
\begin{equation}
Y_{3/2} = Y_\phi / N = \frac{1}{N} \, \sqrt{\frac{m_\phi}{M_P}} \,\,.
\label{abnos}
\end{equation}
If we now take $N \sim 100$ as suggested by the degrees of freedom of the Minimal Supersymmetric Standard Model, the bound~(\ref{limit}) requires
\begin{equation}
m_\phi \lsim 10^9 \, {\rm GeV} \,\,.
\label{fbound2}
\end{equation}
We see that this constraint is much stronger then the one coming from the thermal production~(\ref{fbound1}).

Of course, if it happens that the decay inflaton $\rightarrow$ inflatino
$+$ gravitino is kinematically forbidden $(\vert m_\phi - m_{\tilde \phi}
\vert < m_{3/2}$), then the bound~(\ref{fbound2}) does not
hold.~\footnote{For the specific choice of a present supersymmetry
breaking provided by a Polonyi superpotential~(\ref{sup}) this decay
channel is indeed kinematically forbidden, since one has~\cite{tre}
\begin{equation}
\vert m_\phi - m_{\tilde \phi} \vert \simeq \left( 1 - \sqrt{3}/2 \right) m_{3/2}\, < m_{3/2} \,\,.
\end{equation}
Anyhow, such a relation need not be valid in general.} In addition, if
the scale of supersymmetry breaking is significantly below that of
inflation, i.e., $\mu \ll \Delta$, then even though the decay $\phi \to
{\tilde \phi} + {\tilde G}$ is allowed, it will be naturally kinematically
suppressed with respect to the other decay channels of the inflaton field.
This will open an allowed window for $m_\phi$ even in the simplest
scenarios. Independent of the details of the decay, the rate will always
carry a final state momentum suppression factor.  The overall decay rate
can be written as $\Gamma \sim (1/m_\phi) |{\cal M}|^2 (p/m_\phi)$, where
$|{\cal M}|$ is the amplitude for decay and the final state momentum
suppression factor is
\begin{equation}
2p/m_\phi = \left( 1 - \frac{2(m_{\tilde \phi}^2 + m_{3/2}^2)}{m_\phi^2} +
\frac{(m_{\tilde \phi}^2 - m_{3/2}^2)^2}{m_\phi^4}   \right)^{1/2}
\sim \frac{m_{3/2}}{m_\phi}
\label{poverm}
\end{equation}
Thus in models in which $m_{3/2} \ll m_\phi$, there will be a significant
suppression in the production of gravitinos by either inflaton or
inflatino decay (note that additional suppression may come from the
specific form of the amplitude ${\cal M}$\, as well).

If we take into account the suppression factor~(\ref{poverm}) in the bound
for the direct gravitino production by inflaton decay, and we combine it
with the limit coming from the thermal production, we find that the inflaton mass has to lay within the interval~\cite{tre}
\begin{equation}
10^8 \, {\rm GeV} \, \left( \frac{m_{3/2}}{100 \, {\rm GeV}} \right)^2
\lsim m_\phi \lsim 10^{12} \, {\rm GeV} \,\,.
\label{range}
\end{equation}
Note that the upper bound on $m_\phi$ comes from avoiding a too quick
inflaton decay, with a consequent too high reheating temperature. The
lower bound is instead due to the fact that for too low inflaton mass the
kinematical suppression factor~(\ref{poverm}) is no longer capable of
maintaining a safe small branching ratio for the $\phi \rightarrow {\tilde
\phi} + {\tilde G}$ channel.

The simplest models of single scale inflation with purely gravitational
decays and a scale set by~(\ref{cobeve}), have typically $F \simeq M_P$
and $\Delta \simeq {\rm few} \times 10^{-\,4} \, M_p\,$. In this case the
gravitino mass is slightly higher than $10^{10} \,$ GeV, well within the
allowed range~(\ref{range}). We thus conclude that these models do not
suffer from a gravitino problem.

\Acknowledgements
This work is supported by the European Commission RTN programmes HPRN-CT-2000-00131, 00148, and 00152. We thank Keith~A.~Olive and Lorenzo~Sorbo for an enjoyable collaboration.

\end{document}